\documentclass[aip,reprint]{revtex4-1}

\usepackage[utf8]{inputenc}
\usepackage[T1]{fontenc}

\usepackage[english]{babel}
\usepackage{graphicx}
\usepackage{amsmath}
\usepackage{amssymb}
\usepackage{hyperref}
\usepackage{xcolor}
\usepackage{stmaryrd}

\newcommand {\e} {\varepsilon}
\newcommand {\vp} {\varphi}
\def\const{\mbox{const}}
\def\w{\omega}
\def\W{\Omega}

\newcommand{\ii}{\mathrm{i}}
\newcommand{\dd}{\,\mathrm{d}}

\begin{document}

\title{Desynchronizing two oscillators while stimulating and observing only one}
\author{Erik T.K. Mau}
\email[]{erikmau@uni-potsdam.de}
\affiliation{Department of Physics and Astronomy, University of Potsdam, 
Karl-Liebknecht-Str. 24/25, D-14476 Potsdam-Golm, Germany}

\author{Michael Rosenblum}
\email[]{mros@uni-potsdam.de}
\affiliation{Department of Physics and Astronomy, University of Potsdam, 
Karl-Liebknecht-Str. 24/25, D-14476 Potsdam-Golm, Germany}

\date{\today}

\begin{abstract}
Synchronization of two or more self-sustained oscillators is a well-known and studied phenomenon, appearing both in natural and designed systems. In some cases, the synchronized state is undesired, and the aim is to destroy synchrony by external intervention. In this paper, we focus on desynchronizing two self-sustained oscillators by short pulses delivered to the system in a phase-specific manner. We analyze a non-trivial case when we cannot access both oscillators but stimulate only one. The following restriction is that we can monitor only one unit, be it a stimulated or non-stimulated one.
First, we use a system of two coupled Rayleigh oscillators to demonstrate how a loss of synchrony can be induced by stimulating a unit once per period at a specific phase and detected by observing consecutive inter-pulse durations. Next, we exploit the phase approximation to develop a rigorous theory formulating the problem in terms of a map. We derive exact expressions for the phase -- isostable coordinates of this coupled system and show a relation between the phase and isostable response curves to the phase response curve of the uncoupled oscillator. Finally, we demonstrate how to obtain phase response information from the system using time series and discuss the differences between observing the stimulated and unstimulated oscillator.
\end{abstract}

\maketitle

\begin{quotation}
Synchronization is a natural phenomenon observed when oscillators interact. In some circumstances, a synchronized state is undesired or even harmful. In recent decades, much research has been conducted to develop open and closed-loop control techniques to control synchrony in a system by external intervention.
This paper focuses on a special example motivated by a neuroscience application. We treat two coupled oscillators with a restriction that stimulation does only influence one of them directly. Another constraint is that we have observational access to only one unit, the stimulated or the other. Our objective is to destroy synchrony by pulsatile stimulation, and we achieve this goal by delivering pulses each time the observed oscillator attains a pre-selected trigger phase.
We demonstrate how to recognize a desired desynchronized state in practice by observing the elapsed time between consecutive phase-triggered pulses. Based on the assumptions of weakly coupled phase oscillators and short pulses, we develop a theoretical framework to describe the system's dynamics in response to this stimulation protocol in terms of a dynamical map. This formulation utilizes the phase-isostable description of oscillatory dynamics. We use that to derive a relation between the response curves of the individual oscillator and the coupled system. Our theoretical results are supported by direct numerical simulations of an example system with coupling functions containing higher harmonic terms.
We discuss the approach's optimization for monitoring the stimulated and the unstimulated oscillator. Subject to optimization is the choice of a proper trigger phase and the strength and polarity of the pulses. We demonstrate how to extract the required information from observations of the system and highlight the approach's limitations.
\end{quotation}

\section{Introduction}

Synchronization of oscillatory sources can be beneficial or harmful. 
Examples of the desired synchrony are power grids' functioning~\cite{arenas2008, motter2013, menck2014, auer2017} and atrial pacemaker cells' coordinated activity~\cite{winfree1980, jalife1984}. On the contrary, Parkinson's disease and epilepsy are often related to an adverse effect of synchrony in large neuronal populations~\cite{lehnertz1995, tass1999, stam2005, little2013, tinkhauser2018}. Numerous model studies suggested various techniques for the control of synchrony to cope with this adverse effect ~\cite{tass2001, tass2001a, rosenblum2004, popovych2005, popovych2006, tukhlina2007, wilson2011, franci2012, lin2013, popovych2017, zhou2017, rosenblum2020, toth2022}. 
These studies exploited models of (infinitely) many or several~\cite{tamasevicius2015} mean-field coupled limit-cycle oscillators and assumed that the control input affects the whole population or at least its significant part(s). The feedback techniques relied on observing the collective dynamics. A general approach called synchronization engineering~\cite{kiss2007, kiss2018} also implies access to all network units.

Here, we consider a particular control problem and propose a method to desynchronize two limit-cycle oscillators. Our study is motivated by a neuroscience problem formulated by Azodi-Avval and Gharabaghi~\cite{Azodi-Gharabaghi-15}, who modeled the effect of phase-specific neuromodulation by deep brain stimulation on the synchronized activity of two brain areas. Treating these areas as macroscopical oscillators, they assumed that measurements from both oscillators were available and exploited the technique from Ref.~\cite{kralemann2013} to determine the phase response curve (PRC) for one of the units. Knowledge of the PRC allows stimulation at the most sensitive phase and thus provides a way to efficient desynchronization; however, the PRC obtained from observation of two interacting units generally differs from the phase response to external stimulation.
As another relevant and motivating application, we mention studies of circadian rhythms using the so-called forced desynchrony protocol~\cite{czeisler1999}.
For example,  de la Iglesia et al.~\cite{delaiglesia2004} exposed rats to an artificial light-dark rhythm with a period of $22$ hours and found that the rats' activity pattern split into the entrained rhythm and another one with a period significantly larger than $24$ hours. This splitting may indicate an enforced desynchronization of individual circadian oscillators.

We elaborate on the idea by Azodi-Avval and Gharabaghi~\cite{Azodi-Gharabaghi-15} and suggest a minimal setup where we achieve desynchronization by observing and perturbing {\it only one unit}. We consider two versions of the approach, where we monitor either the stimulated oscillator or the other. Having in mind a possible neuroscience application, we exploit a pulsatile perturbation delivered approximately once per oscillatory cycle. We remark that models of two coupled phase oscillators with open-loop pulsatile stimulation have been studied in Refs.~\cite{tass2003a, tass2004, krachkovskyi2006}. We also mention that Montaseri et al.~\cite{montaseri2011, montaseri2014} used a feedback controller design inspired by the role of astrocytes in neural information processing to desynchronize two oscillators. However, 
Refs.~\cite{montaseri2011, montaseri2014}
assumed that both systems could be observed and stimulated. 

Finally,  we recall that  Pyragas et al.~\cite{pyragas2007} and Tukhlina et al.~\cite{tukhlina2008} considered synchrony suppression in a model of two interacting oscillator populations, one used for sensing and another for stimulation. Those models can be treated as two coupled macroscopic oscillators.
Furthermore, Hauptmann et al.~\cite{hauptmann2005} considered two unidirectionally coupled oscillatory populations with variable sites for sensing and spatially coordinated stimulation, and Popovych et al.~\cite{popovych2010} considered two interacting populations, one oscillatory and one in equilibrium without coupling, where stimulation entered the oscillatory one only. Both approaches successfully desynchronized the entire system by delayed feedback.
However, desynchronization on the level of subpopulations means quenching of macroscopic oscillators, while our study aims to keep systems oscillating but destroy their synchrony.

This article is structured as follows: First, we illustrate the problem formulation and the detection of stimulation-induced desynchronization using two coupled Rayleigh oscillators in Section~\ref{Sec:II}.
In Section~\ref{sec:theory}, we develop a theoretical framework for two weakly coupled oscillators, describing phase-specific stimulation of the system in terms of a dynamical map. Our theoretical analysis exploits the phase -- isostable representation of oscillatory dynamics~\cite{wilson2016}. Section~\ref{sec:response} shows a relation between the phase and isostable response curves of the synchronized oscillatory dynamic and the phase response curve of an uncoupled oscillator and thus complements the theoretical analysis. Here we also discuss possible approaches to obtain the phase response curve from time series data of only one oscillator. Finally, Section~\ref{sec:discussion} discusses a strategy to optimize the simulation by minimizing the total intervention in the system, as well as open problems and limitations of our approach.

\section{Illustration of the approach}
\label{Sec:II}
\begin{figure}[h!]
\includegraphics[width=0.48\textwidth]{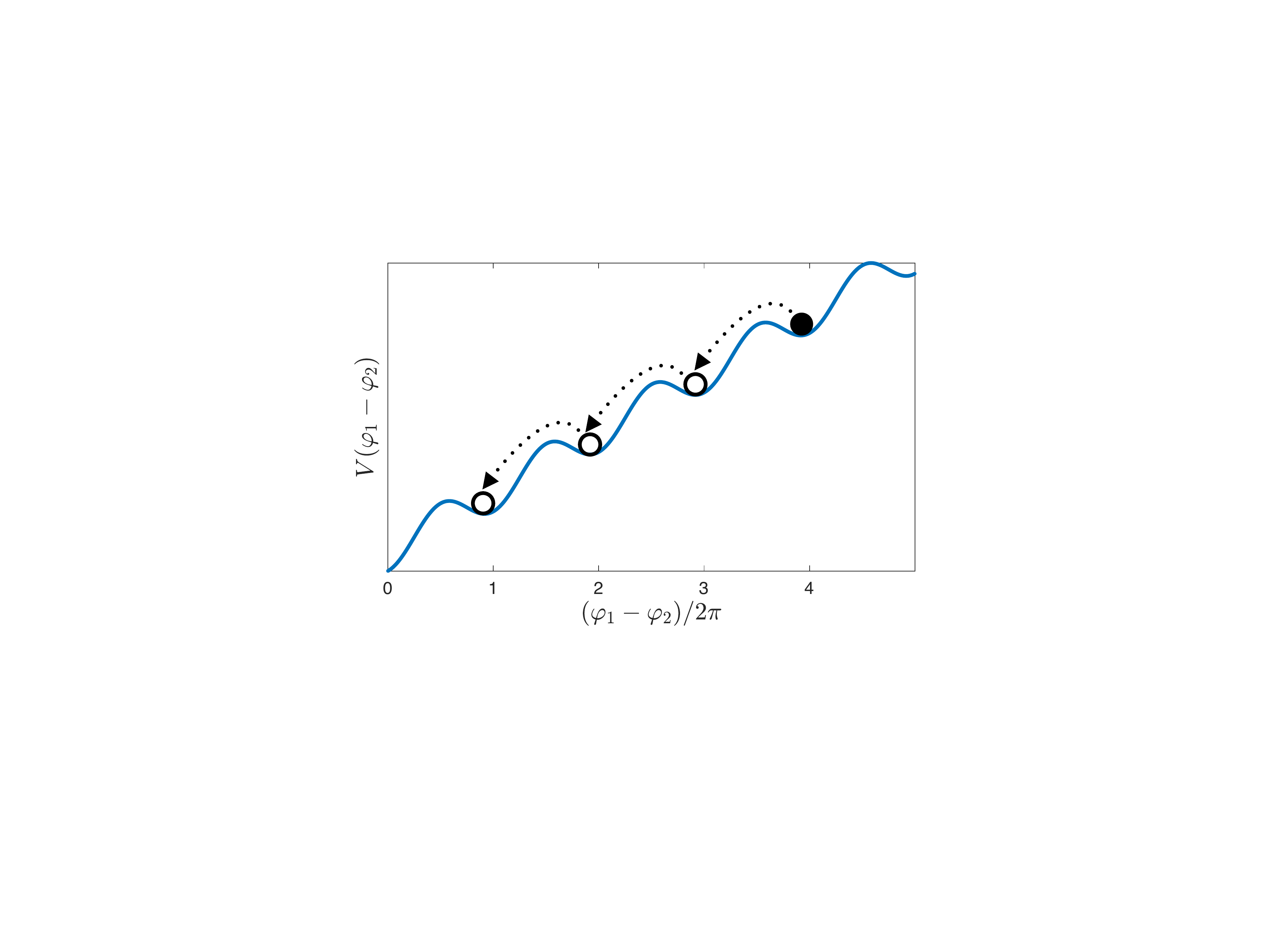}
\caption{The dynamics of the phase difference between two weakly coupled oscillators can be illustrated by the motion of an overdamped particle in an inclined potential, plotted here for the case $\w_1< \w_2$. A synchronous state corresponds to a particle trapped in a minimum of the potential. Stimuli applied at a proper phase can shift the particle from the equilibrium position and eventually move it to the next potential well, decreasing the phase difference $\vp_1-\vp_2$ by $2\pi$, i.e., inducing a phase slip. We aim to design a stimulation that permanently causes phase slips, thus destroying synchrony.
}
\label{fig:potential}
\end{figure}

The general theory says that phase dynamics of two weakly coupled limit-cycle oscillators can be illustrated by the motion of an overdamped particle in an inclined potential, see, e.g.,~\cite{pikovsky2001} and Fig.~\ref{fig:potential}. The particle at rest in a potential well corresponds to the synchronous state with the phase difference $\vp_1-\vp_2=\const$.
Thus, the desynchronization problem reduces to kicking the particle down the potential, inducing phase slips, i.e.,  relatively rapid jumps where the phase difference changes by $\pm 2\pi$.
(Certainly, one can kick the particle to move it up, but this action requires stronger stimulation and, therefore, is less efficient.) 
For that purpose, we consider relatively rare pulses applied approximately once per oscillation period.
Suppose each pulse shifts the particle toward the local maximum. Between two consecutive stimuli, the particle tends to return to equilibrium. This consideration shows that there shall be a critical value of the pulse strength such that the phase shifts accumulate and the particle eventually moves from its stable equilibrium position over the maximum to the following equilibrium position. This way, the phase difference changes by $2\pi$ (phase slip). The continuing stimulation evokes the next phase slip, and so on.

We demonstrate the approach exploiting the system of two coupled Rayleigh oscillators perturbed by a pulse stimulation:
\begin{align}
    \ddot x_{1} & - \mu(1-\dot x_{1}^2)\dot x_{1}+\w_{1}^2 x_{1}
    =\e(x_{2}-x_{1}) + p(t) \;, \\
    \ddot x_{2} & - \mu(1-\dot x_{2}^2)\dot x_{2}+\w_{2}^2 x_{2}
    =\e(x_{1}-x_{2}) \;.
    \label{eq:coupled_Rayleighs}
\end{align}
Parameters are $\mu=2$, $\w_1=0.98$, $\w_2=1.02$, $\e=0.2$.
The perturbation $p(t)$ is a pulse train, $p(t)=\sum_k {\cal P}(t_n)$, where ${\cal P}(t_n)$ is a finite-length pulse applied at the instant $t_n$.
We note that we label the stimulated unit as the first for definiteness.
Next, without loss of generality, we choose $\w_1 < \w_2$; to treat the opposite choice $\w_1 > \w_2$, one has to choose another stimulation phase, as discussed below.  

We now discuss the determination of the stimulation times $t_n$. Suppose we observe $x_1(t)$. We define threshold-crossing events $t_n$ as the instants when $x_1(t_n)=x_0$ and $\dot x_1(t_n)$ is either always positive or always negative; here, $x_0$ is the threshold value. (The proper choice of $x_0$ and condition for $\dot x_1$ is discussed below in Section~\ref{sec:response}.) We apply pulses at $t_n$ with the following additional restriction. Suppose for definiteness that we choose the condition $\dot x_1>0$. If the pulse applied at $t_n$ reduces $x_1(t)$, then after a very short time interval $\Delta t \ll T_0$, where $T_0$ is the period of synchronous oscillation, $x_1(t)$ again achieves the threshold value $x_0$. We neglect this threshold crossing and wait till the next one so that the intervals $\tau_n=t_{n+1}-t_n$ are of the order of $T_0$. We denote the return times $\tau_n$ as partial periods of the first oscillator. The formulated condition can be easily explained in terms of the oscillator's phase. Indeed, the threshold condition $x_1(t_n)=x_0$ corresponds to achieving a certain phase $\vp_0$. 
Stimulation can decrease $\vp_0$; thus, for the subsequent stimulation, we wait until the oscillator's phase becomes $\vp_0+2\pi$. 
A similar consideration applies when we monitor $x_2$.

\begin{figure}[!ht]
\includegraphics[width=0.48\textwidth]{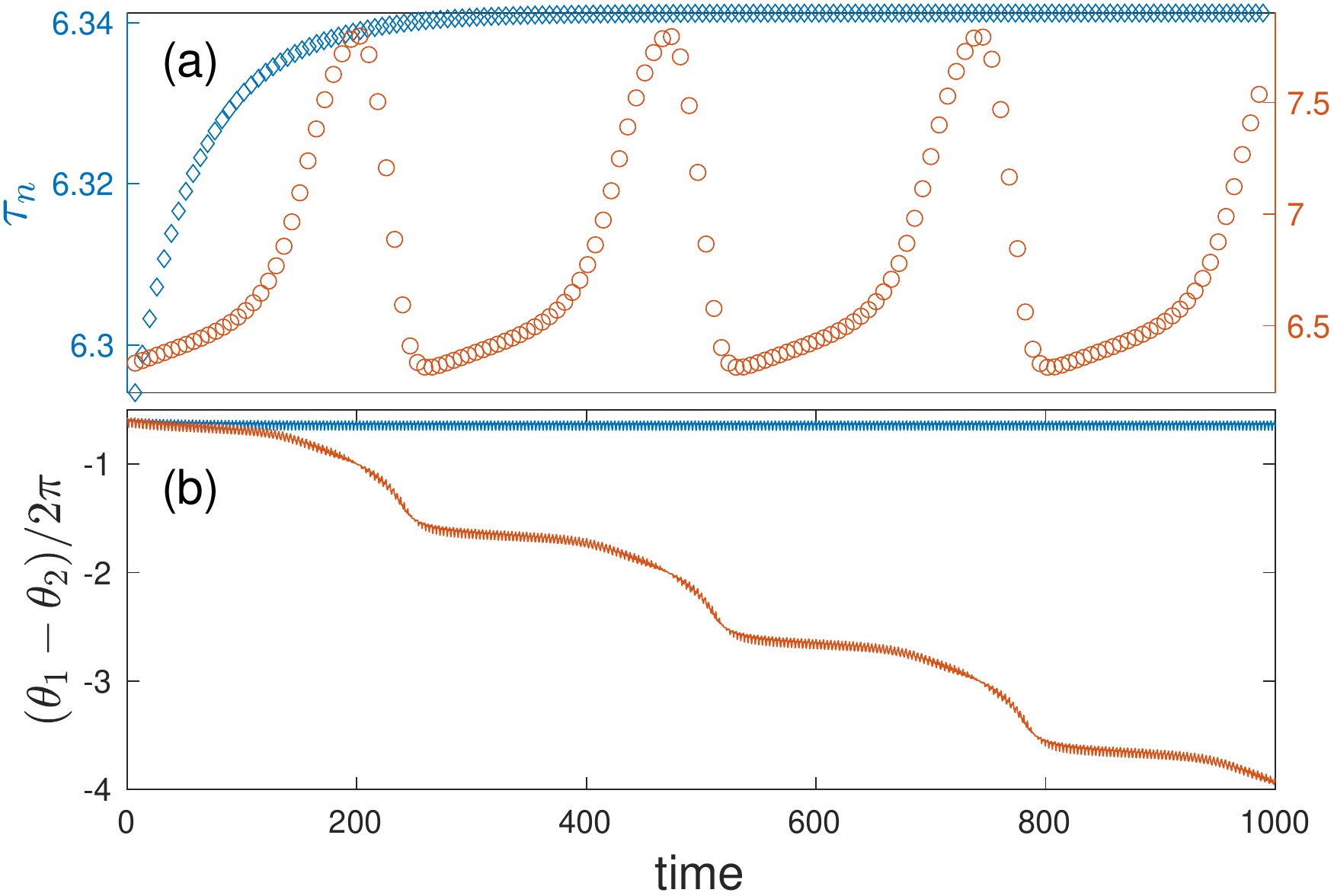}
\caption{(a) Partial periods of the stimulated Rayleigh oscillator vs. stimulation times, for weak, $I=2$, and strong, $I=4$ stimulation (blue diamonds, left vertical axis, and red circles, right vertical axis, respectively). In both cases, the stimulation changes the period. However, when the stimulation amplitude is below a certain threshold, the two coupled oscillators remain synchronized, as seen from the 
(proto)phase difference depicted in (b). If the stimulation is sufficiently strong, it induces phase slips; the occurrence of phase slips can be traced from the oscillations of $\tau_n$.
}
\label{fig:ray_effect}
\end{figure}

We illustrate the effect of stimulation by plotting the partial periods $\tau_n$ vs. $t_n$ in Fig.~\ref{fig:ray_effect}a, for $x_0=1$, $\dot x_1>0$. Panel (b) shows the protophase difference~\footnote{In this plot, we operate with $\theta$, which is the polar angle in the $x,\dot x$ plane. This variable (protophase) differs from the true phase on the time scale of a single period; this difference is not essential here since we are interested in the presence or absence of phase slips.}.
We use rectangular pulses of duration $\Delta = 0.01$ and amplitude $I$.
Inspecting the plot, we conclude that oscillation of $\tau_n$ indicates phase slips and, hence, a desynchronizing action.~\footnote{
Destruction of synchrony, i.e., a transition from periodic to quasiperiodic motion, can be traced in the power spectrum of $x_1$. However, achieving the required spectral resolution requires a relatively long time series.}

\begin{figure}[!ht]
\includegraphics[width=0.48\textwidth]{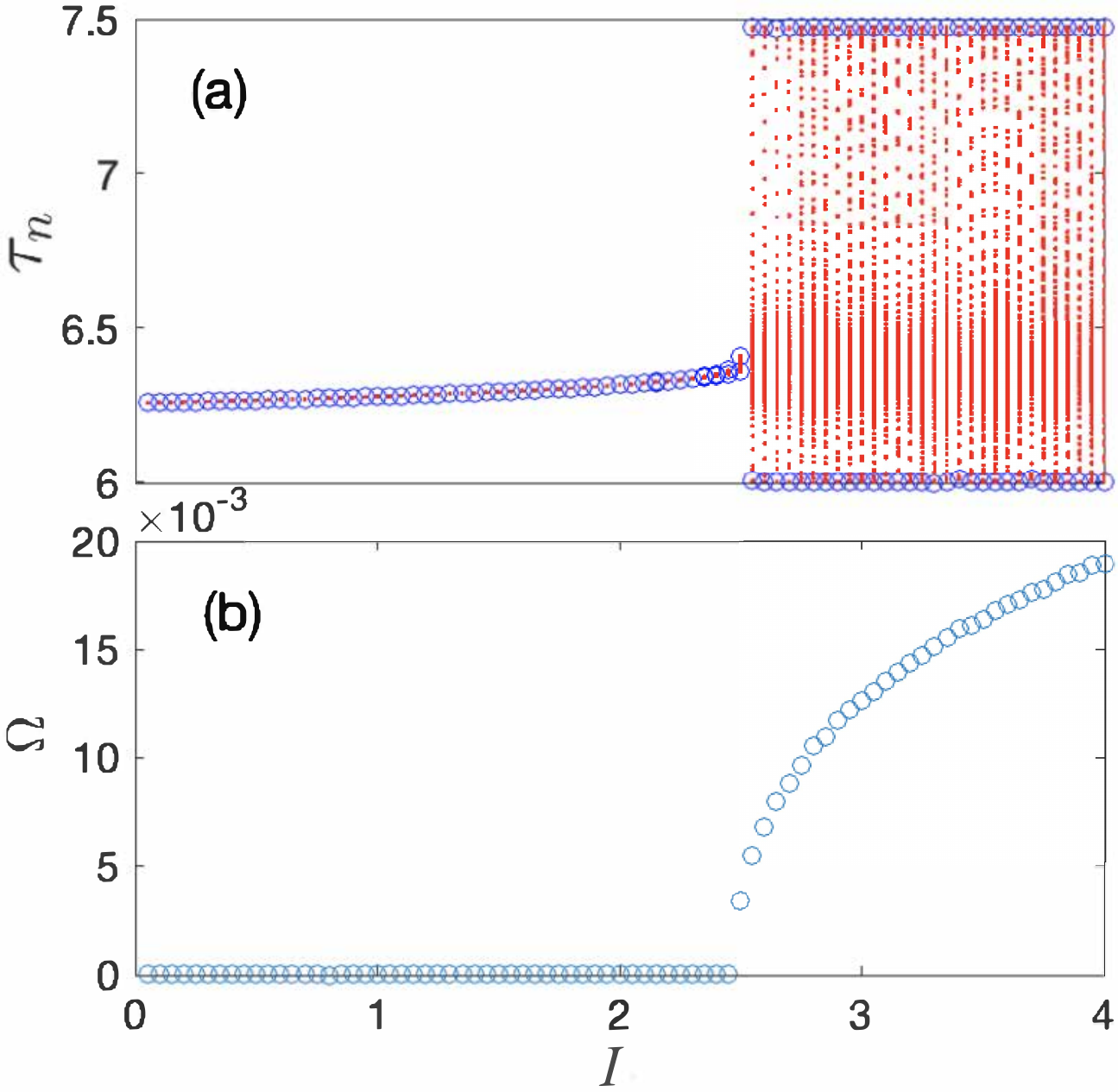}
\caption{Illustration of the case when the first Rayleigh oscillator is stimulated by a pulse whenever the phase of the second one attains a specific fixed value. In (a), we plot values of the second unit periods $\tau_n$ vs. the stimulation amplitude $I$ (red dots); for better visibility, we also show the minimal and maximal values of $\tau_n$ for each $I$ (blue circles). For $I\lesssim 2.55$ we have $\tau_{min}=\tau_{max}$, what means that the system remains synchronized. For $I\gtrsim 2.55$, the partial periods $\tau_n$ oscillate, indicating phase slips and loss of synchrony. The loss of synchrony is confirmed in panel (b), where we demonstrate the difference $\W$ of oscillator frequencies that is non-zero for $I\gtrsim 2.55$.}
\label{fig:ray_effect2}
\end{figure}

Figure~\ref{fig:ray_effect2} depicts the case when we observe the second oscillator. Thus, we define the partial periods, now for the second oscillator, as $\tau_n=t_{n+1}-t_n$ via the events $t_k$ when $x_2(t_n)$ crosses a certain threshold, e.g., in the positive direction. Omitting the first 50 intervals, we plot $\tau_n$, $\tau_{min}=\text{min}(\tau_n)$, and $\tau_{max}=\text{max}(\tau_n)$, $n>50$, for different values of the pulse amplitude $I$. We used $x_0=-1$; other parameters are the same as in Fig.~\ref{fig:ray_effect}. We see that sufficiently strong stimulation results in oscillatory behavior of $\tau_n$, which means the appearance of phase slips, and, hence, desynchronization.
Thus, we can desynchronize the system by stimulating only one of two synchronous oscillators while observing any of these two. We support this conclusion with theoretical analysis in the next Section.

\section{Desynchronizing by pulse stimulation: theory}
\label{sec:theory}

It is well-known that, for sufficiently weak coupling, phase dynamics of two interacting units obey the Kuramoto-Daido equations:
\begin{align}
    \dot{\vp}_1 &= \w_1 + C_1(\vp_1 - \vp_2) + Z(\vp_1)p(t)\;, \label{phmodel1}\\
    \dot{\vp}_2 &= \w_2 + C_2(\vp_2 - \vp_1)\,,   \label{phmodel2}
\end{align}
where $C_{1,2}$ are coupling functions.
Here, we assume for definiteness that $\w_1<\w_2$ and that stimulation $p(t)$ affects the first oscillator. 
The last term in Eq.~(\ref{phmodel1}) describes the stimulation, where $p(t)$ is the external force, and the phase response curve (PRC) of the uncoupled oscillator $Z(\vp_1)$ quantifies the sensitivity of the unit to perturbation. 
We will consider separately two cases where we observe either the first or the second oscillator. 
Therefore, introducing the phase difference $\eta=\vp_1-\vp_2$ we re-write Eqs.~(\ref{phmodel1},\ref{phmodel2}) as equations for $\eta,\vp_i$, where either $i=1$ or $i=2$: 
\begin{align}
    \dot\eta & =f(\eta)+ Z(\vp_i+\delta_{i2}\eta)p(t)  
    \label{phmodel3}\;,\\
    \dot{\vp_i} &= g_i(\eta) + \delta_{i1}Z(\vp_i+\delta_{i2}\eta)p(t)
    \label{phmodel4}\,.
\end{align}
Here, $\delta_{i2}$ is the Kronecker symbol, $g_1(\eta)=\w_1+C_1(\eta)$,  $g_2(\eta)=\w_2+C_2(-\eta)$, and $f(\eta)=g_1(\eta)-g_2(\eta)$.
Note that PRC $Z$ remains the function of $\vp_1 = \vp_i+\delta_{i2}\eta$.

Suppose there are no perturbations, $p(t)=0$.
Then, Eq.~(\ref{phmodel3}) reduces to $\dot\eta=f(\eta)$.
The dynamics of this equation are well-studied. Depending on the parameters, it has either asynchronous solution $\dot\eta<0$ or synchronous, phase-locked solution $\dot\eta=0$.
In the latter case, one or several pairs of stable and unstable fixed points exist. 
We present the theory for the case when there exists only one stable fixed point $\eta^*=\const$, $f'(\eta^*)<0$, and discuss a possible extension to the general case in Section~\ref{sec:discussion}.
Asynchronous solutions correspond to quasiperiodic trajectories on the two-torus spanned by $\vp_1,\vp_2$. In contrast, the existence of stable and unstable fixed points in Eq.~(\ref{phmodel3}) means the appearance of stable and unstable limit cycles on the torus.

Consider the stable limit cycle on the two-torus. The frequency of this synchronous solution is $\w=g_i(\eta^*)$. Next, we define the phase on the limit cycle and in its vicinity. We emphasize that the phase of the uncoupled oscillator $\vp_i$ is not the true asymptotic phase of the synchronous solution of the coupled system because its time derivative is not a constant but depends on $\eta$, see Eq.~(\ref{phmodel4}). Thus, in the context of the coupled system, we treat $\vp_i$ as the protophase (angle variable). Using the ansatz $\Phi(\vp_i,\eta)=\vp_i+ \delta_{i2}\eta^* + F_i(\eta)$ with an additional condition $F_i(\eta^*)=0$, we require $\dot \Phi=\w$ and obtain
\[
\dot \Phi(\vp_i,\eta)=\dot\vp_i + F'_i(\eta)\dot\eta=g_i(\eta)+F'_i(\eta)f(\eta)=\w\;.
\]
Solving this equation for $F'_i$ and integrating, we obtain~\footnote{For the actual computation of $\Phi$ it is important to avoid integrating over a singularity $f(s)=0$, e.g., at the unstable phase difference.}: 
\begin{equation}
\Phi(\vp_i,\eta)=\vp_i+\delta_{i2}\eta^*+\int_{\eta^*}^\eta\frac{\w-g_i(s)}{f(s)}\dd s \;.
\label{eq:phase_general}
\end{equation}
Using $f(\eta)=g_1(\eta)-g_2(\eta)$, it is easy to check that $\Phi(\vp_1,\eta) = \Phi(\vp_2,\eta)$, i.e., the definition of phase does not depend on the chosen protophase. On the limit cycle ($\eta=\eta^*$), we have $\Phi = \vp_1 = \vp_2 + \eta^*$, meaning phase and protophase coincide up to a constant shift. We remind that by construction, $\Phi(\vp_i,\eta^*)=\vp_1$, i.e. the protophase $\vp_1$ coincides with $\Phi$ on the limit cycle.

Before proceeding with a separate analysis of the cases $i=1$ (the first oscillator is observed) and $i=2$ (the second unit is observed), we conclude the theoretical consideration by the following remark. For the attractive cycle on the torus, 
$\eta-\eta^*=\psi$ describes the deviation from the stable solution; hence, the variable $\eta$ plays the role of the amplitude. In a small vicinity of the limit cycle~\footnote{We note that one can adequately introduce the isostable variable as $\psi:=\frac{f(\eta)}{\kappa}\exp \left(\int_{\eta^*}^\eta \frac{\kappa - f'(s)}{f(s)} \dd s \right) $ so that equation $\dot\psi=\kappa\psi$ is valid in the whole basin of attraction of the limit cycle. Introduced in this way, $\psi$ generally 
differs from $\eta-\eta^*$ if the quantity is not small.}, we then write $\dot\psi=f'(\eta^*)\psi=\kappa\psi$ and interpret $\psi$ as the isostable variable~\cite{wilson2016}. 
We return to the phase -- isostable representation of the synchronized dynamics in Section~\ref{sec:response}. 

\subsection{Stimulating and observing the same oscillator}

Here, we assume we observe the first unit and compute the intervals between the stimuli. We recall that we stimulate each time the phase of the first oscillator attains some fixed value $\vp_0$.
Let the variable $\eta$ immediately before the $n$-th stimulus is $\eta_n$.
We assume instantaneous phase shift due to the $\delta$-kick, i.e., ${\cal P}(t_n)=q\delta(t-t_n)$, so that $\vp_1=\vp_0\to\vp_0+A$ and $\eta\to\eta+A$, where the instantaneous phase shift $A=qZ(\vp_0)$ and $q$ is the amplitude of the $\delta$-pulse. As before, we denote the time between the $n$-th and $n+1$-th kick by $\tau_n$. Between the stimuli, we deal with autonomous dynamics.
Hence, $\tau_n$ is obtained by
\begin{align}
    \tau_n = \int_{\eta_n+A}^{\eta_{n+1}} \frac{\dd s}{f(s)} 
    \label{eq:tau_n_general}
    \,,
\end{align}
and the phase $\Phi$ within this time interval grows by $\omega \tau_n$. We thus write
\begin{equation}
    \Phi(\vp_0+A,\eta_n+A)+\w\tau_n=\Phi(\vp_0+2\pi,\eta_{n+1})\;.
    \label{eq:phase_growth}
\end{equation}
Exploiting the definition of phase from Eq.~\eqref{eq:phase_general}, 
we obtain the equation
\begin{align}
    A + \int^{\eta_n+A}_{\eta^*} \frac{\w-g_1(s)}{f(s)} \dd s + \w \tau_n = 2\pi + \int^{\eta_{n+1}}_{\eta^*} \frac{\w-g_1(s)}{f(s)} \dd s 
    \label{eq:map_intermediate_step}
    \,.
\end{align}
By inserting the expression of $\tau_n$ from Eq.~\eqref{eq:tau_n_general} into this formula, we finally obtain
\begin{align}
    2\pi - A - \int_{\eta_n+A}^{\eta_{n+1}} \frac{g_1(s)}{f(s)} \dd s = 0
    \label{eq:map_definition}
    \,.
\end{align}
This equation defines a one-dimensional map $\eta_{n+1} = \mathcal{F}(\eta_n)$ with the parameter $A$. We iterate this map, starting from $\eta_0 = \eta^*$ and solving Eq.~\eqref{eq:map_definition} numerically~\footnote{From SciPy~\cite{virtanen2020}, we exploit the integration algorithm \textit{scipy.integrate.quad} and root-finding algorithm \textit{scipy.optimize.root} with solver method \textit{hybr} (modified Powell hybrid method). We define the r.h.s. of Eq.~\eqref{eq:map_definition} as a function of $\eta_{n+1}$ with parameters $A$ and $\eta_n$. We obtain $\eta_{n+1}=\mathcal{F}_A(\eta_n)$ by calling the root-finding on this function with initial guess $\eta_n+A$. For faster computation, we circumvent to execute the root-finding algorithm every time we call $\mathcal{F}_A$ by computing $\mathcal{F}_A(\eta)$ for a sufficiently large set of $\eta$-values once and fitting this to a finite Fourier series.}, for a fixed kick strength $A$. Using the obtained values of $\eta_n$, we integrate numerically Eq.~\eqref{eq:tau_n_general} and obtain $\tau_n$.
We remind that the sequence of intervals $\tau_n$ can easily be measured in an experiment.

In Section~\ref{Sec:II}, we have demonstrated that depending on the stimulation strength, the sequence $\tau_n$ either saturates or oscillates, see Fig.~\ref{fig:ray_effect}. The former case means that the map $\eta_{n+1} = \mathcal{F}(\eta_n)$ has a fixed point $\hat\eta(A)$ with an obvious condition $\hat\eta(0)=\eta^*$. We denote the corresponding interval $\hat\tau(A)$, where $\hat\tau(0)=2\pi/\w$.

For small $A$ both $\eta_n+A$ and $\eta_{n+1}$ are close to $\eta^*$ and we can write the first-order approximation of Eq.~\eqref{eq:map_intermediate_step}. For this purpose, we use
$\w=g_1(\eta^*)$ and compute $\lim_{\eta\to\eta^*}(\w-g_1(s))/f(s)$ using the L'Hospital's rule. We obtain:
\begin{align}
    A-\frac{g_1'(\eta^*)}{f'(\eta^*)}(\eta_n+A-\eta^*) + \w\tau_n
    = 2\pi-\frac{g_1'(\eta^*)}{f'(\eta^*)}(\eta_{n+1}-\eta^*)
    \;.
    \label{eq:map_intermediate_step_at_LC_osc1}
\end{align}
In the following, we define $\gamma = 1-g'_1(\eta^*)/f'(\eta^*)$. The approximation~\eqref{eq:map_intermediate_step_at_LC_osc1} yields the intervals $\tau_n$ in the vicinity of $\eta^*$, i.e., for small kick strength $A$ as
\begin{align}
    \tau_n
    &= \frac{1}{\w}[2\pi - \gamma A +(\gamma-1)(\eta_{n+1}-\eta_n)]\;.
\end{align}
Imposing the fixed point condition $\eta_n=\eta_{n+1}$ and inserting $A=qZ(\vp_0)$ we obtain an expression for $\hat\tau$ as
\begin{align}
    \hat\tau = \frac{1}{\w} (2\pi- \gamma qZ(\vp_0))
    \label{eq:kick_tau_fixed_point_osc1}
    \;.
\end{align}
We remark that the direction of convergence to that fixed point depends on the sign of $\gamma-1$. There may be a $\tau_n$ in the transient that is larger or smaller than both $\hat{\tau}$ and $2\pi/\w$.

\subsection{Stimulating the first oscillator while observing the second one}
\label{sec:theory_osc2}

Now, we use the events $\vp_2=\vp_0$ as a trigger for stimulation. Again, we aim to describe the dynamic via a one-dimensional map $\eta_{n+1}={\cal F}(\eta_n)$. 
The effect of the kick is now $\vp_1 \to \vp_1+qZ(\vp_1)=\vp_1+qZ(\vp_0+\eta)$ and, hence, $\eta \to \eta + qZ(\vp_0 + \eta)$.
The evoked shift of $\eta$ depends on $\eta$ itself and is not constant as in the previous case. Thus, we cannot combine the kick action $q$, the trigger phase $\vp_0$, and the response $Z(\vp_0)$ into a  constant phase shift, but have to treat it as a function $qZ(\vp_0+\eta)$ evaluated at $\eta_n$. For convenience, we denote $\Tilde{Z}(\eta) := qZ(\vp_0 + \eta)$. Accordingly, the interval $\tau_n$ between two kicks is 
\begin{align}
    \tau_n = \int_{\eta_n+\Tilde{Z}(\eta_n)}^{\eta_{n+1}} \frac{\dd s}{f(s)} 
    \label{eq:tau_n_general_phi2}
    \,.
\end{align}

Proceeding as in the previous case, we write, similarly to Eq.~\eqref{eq:phase_growth}:
\begin{equation}
    \Phi(\vp_0, \eta_n+\Tilde{Z}(\eta_n))+\w\tau_n=\Phi(\vp_0+2\pi,\eta_{n+1})
    \label{eq:map_intermediate_step_osc2}
    \;.
\end{equation}
Finally, we obtain the equation 
\begin{align}
    2\pi - \int_{\eta_n+\Tilde{Z}(\eta_n)}^{\eta_{n+1}} \frac{g_2(s)}{f(s)} \dd s = 0
    \label{eq:map_definition_osc2}
    \,
\end{align}
that defines the map $\eta_{n+1}={\cal F}(\eta_n)$ depending on function $\Tilde{Z}$.

Similarly to the previous case, we find an approximate expression for  $\tau_n$ in the limit of weak kicks leaving the phase difference close to $\eta^*$. We approximate Eq.~\eqref{eq:map_intermediate_step_osc2} by
\begin{align}
    -\frac{g_2'(\eta^*)}{f'(\eta^*)}(\eta_n + \Tilde{Z}(\eta_n)-\eta^*) + \w\tau_n
    = 2\pi-\frac{g_2'(\eta^*)}{f'(\eta^*)}(\eta_{n+1}-\eta^*)
    \;.
    \label{eq:map_intermediate_step_at_LC_osc2}
\end{align}
Note that $-g_2'(\eta^*)/f'(\eta^*)$ equals the above defined constant $\gamma$ used in the previous case of monitoring the first oscillator. This can be checked by inserting the original coupling functions $C_1$ and $C_2$ into $f, g_1, g_2$. For $\tau_n$ we obtain
\begin{align}
    \tau_n = \frac{1}{\w}(2\pi - \gamma \Tilde{Z}(\eta_n) +\gamma(\eta_{n+1}-\eta_n))
    \,.
\end{align}
In the limit of small $q$ we conclude $\Tilde{Z}(\eta_n) = qZ(\phi_0 + \eta_n)\approx qZ(\phi_0 + \eta^*)$. Thus for the fixed point $\hat{\tau}$, we obtain a result similar to that of the first case:
\begin{align}
    \hat\tau = \frac{1}{\w} (2\pi- \gamma Z(\phi_0 + \eta^*) q)
    \label{eq:kick_tau_fixed_point_osc2}
    \,.
\end{align}
Compared to Eq.~\eqref{eq:kick_tau_fixed_point_osc1}, the only difference is the argument of $Z$. In the case of the first oscillator being monitored, it is $\vp_0$, and in the current case, it is $\vp_0 + \eta^*$. We remark that by the definition of phase via Eq.~(\ref{eq:phase_general}), in both cases we have $\Phi_0 = \vp_0 + \delta_{i2}\eta^*$. Thus, in both cases the expression for $\hat{\tau}$ in the limit of small $q$ reads
\begin{align}
    \hat\tau = \frac{1}{\w} (2\pi- \gamma qZ(\Phi_0))
    \label{eq:kick_tau_fixed_point}
    \;,
\end{align}

In the following, we will test the derived dynamical map $\mathcal{F}$ for a model of coupled phase oscillators and compare it to a direct simulation with both finite-size and Dirac kicks.

\subsection{An example: coupled phase oscillators}
\label{sec:theory_example}

\begin{figure}[!ht]
    \includegraphics[width=0.48\textwidth]{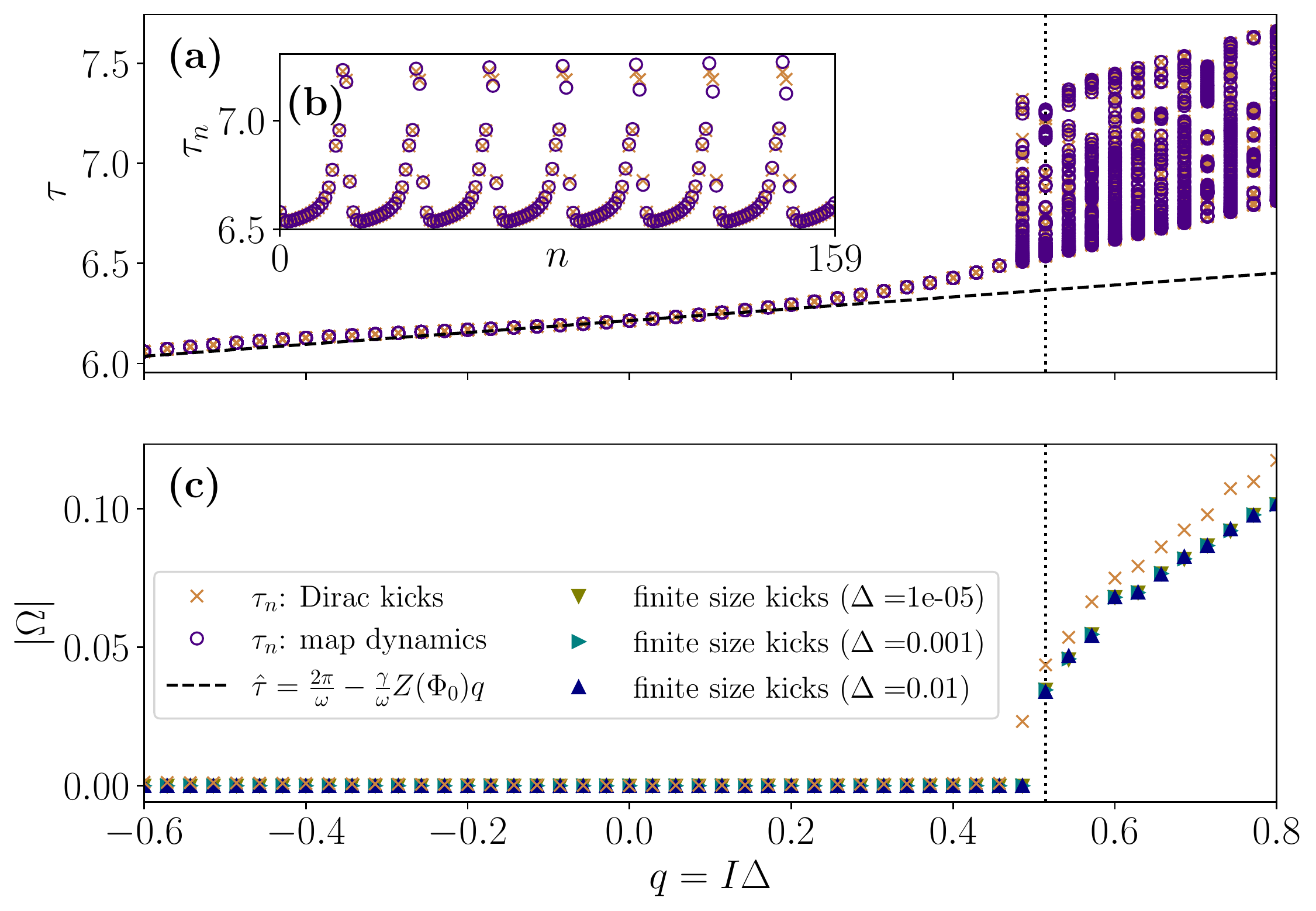}
    \caption{Dynamics of the kicked phase oscillator system~\eqref{eq:phase_oscillator_system_example}. Here, we monitor the first oscillator and deliver kicks at $\Phi_0 = \vp_0 = 3\pi/2$. Panel (a) depicts the bifurcation diagram for the asymptotic behavior of inter-kick intervals $\tau_n$. The values of $\tau_n$ for $n \geq 50$ are shown as a function of the kick action $q$ for a direct simulation of Dirac kicks (orange crosses), and the iteration of the map $\mathcal{F}$ (purple circles). The approximate expression~\eqref{eq:kick_tau_fixed_point} for the fixed point $\hat{\tau}$ is drawn as a black dashed line. For small kick actions, the $\tau_n$ converge to a fixed point in first-order approximation given by $\hat{\tau}$. Phase slips occur for sufficiently large values of $q$. An example in (b), depicts the inter-kick durations $\tau_n$ for $q\approx 0.51$ (this value is marked with a dotted line in (a) and (c)). Panel (c) shows the bifurcation diagram for the time-averaged frequency difference $|\Omega|$ of both oscillators (time averaging over $80$ kicks). Data points correspond to direct simulations of Dirac kicks (orange crosses) and finite-sized kicks with pulse widths $\Delta=10^{-5}$ (olive lower triangles), $\Delta=10^{-3}$ (green right triangles), and  $\Delta=10^{-2}$ (dark blue upper triangles). The emergence of a non-zero value of $|\Omega|$ coincides with the disappearance of the stable fixed point and onset of oscillatory dynamics for $\tau_n$ in (a).}
    \label{fig:bifurcation_diagram_osc1}
\end{figure}

\begin{figure}[!ht]
    \includegraphics[width=0.48\textwidth]{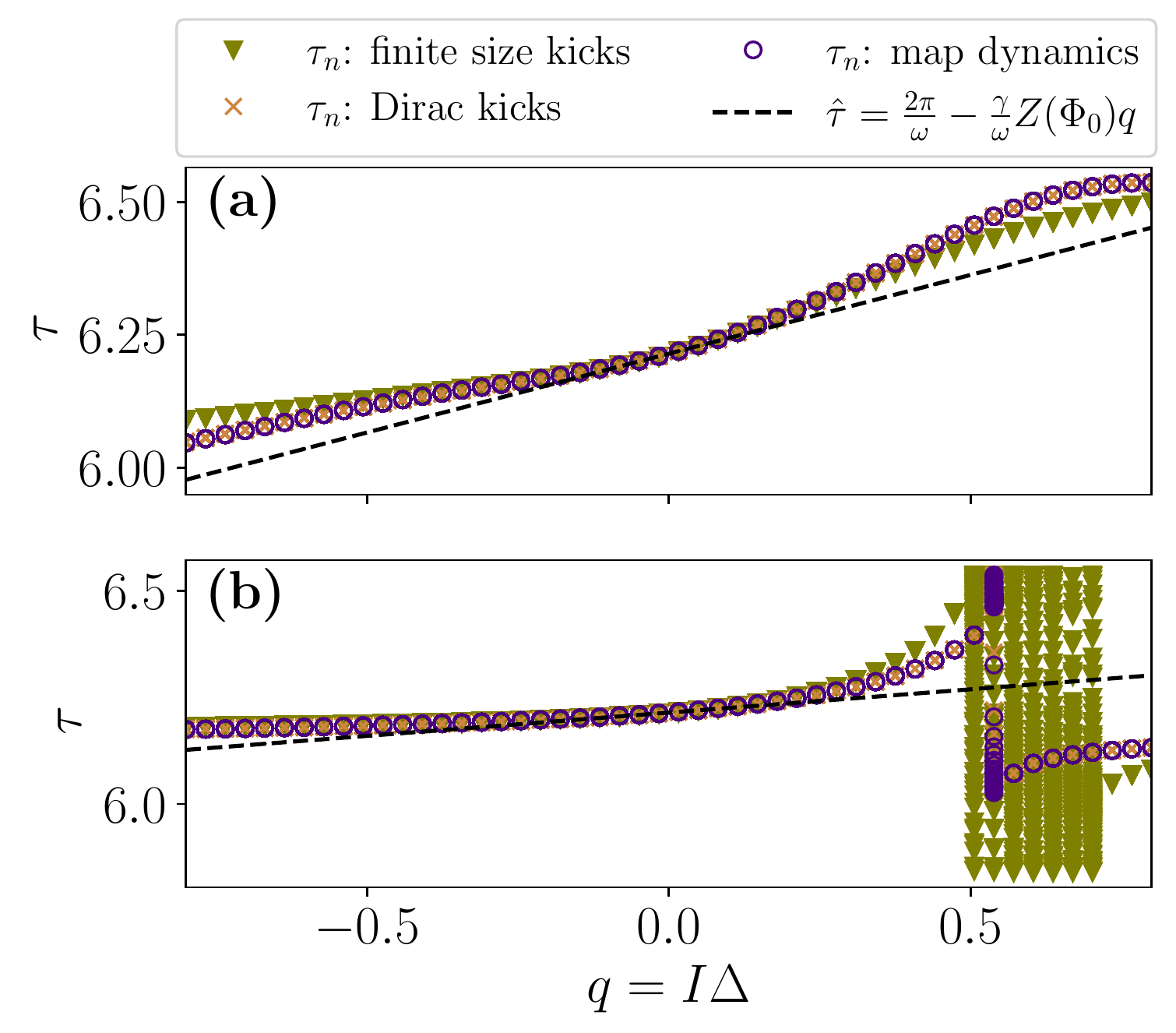}
    \caption{Bifurcation diagrams of the kicked phase oscillator system~\eqref{eq:phase_oscillator_system_example} when monitoring the second oscillator, for the cases of $\Phi_0 = 3\pi/2$ ($\vp_0 \approx 1.62 \pi$, panel (a)) and $\Phi_0 = 1.88 \pi$ ($\vp_0 \approx 0$, panel (b)). The values of $\tau_n$ for $n \geq 50$ are depicted as a function of the kick action $q$ for direct simulation of finite-size kicks (green triangles; pulse duration $\Delta = 10^{-5}$ and amplitude $I= q/\Delta$), Dirac kicks (orange crosses), and the iteration of the map $\mathcal{F}$ (purple circles). The expression~\eqref{eq:kick_tau_fixed_point} for the fixed point $\hat{\tau}$ is drawn as a black dashed line. Phase slips do not occur in (a). In (b) phase slips occur only in an interval of kick actions $q$ which is different for Dirac and finite-sized kicks.}
    \label{fig:bifurcation_diagram_osc2}
\end{figure}

We consider two phase oscillators with coupling functions containing higher harmonics terms
\begin{equation}
\begin{array}{rcl}
    \dot{\vp}_1 &=& \w_1 + \e \sin(\vp_2 - \vp_1) + \sigma \sin(2(\vp_2 - \vp_1)) + Z(\vp_1)p(t)\;,\\
    \dot{\vp}_2 &=& \w_2 + \e \sin(\vp_1 - \vp_2) + \beta \sin(3(\vp_1 - \vp_2))
\label{eq:phase_oscillator_system_example}
    \;,
\end{array}
\end{equation}
with the parameters $\w_1 = 0.98$, $\w_2 = 1.02$, $\varepsilon= 0.05$, $\sigma = 0.02$ and $\beta = -0.01$. For the response curve $Z$, we choose a simple sine function $Z(\vp_1)=\sin(\vp_1)$.

Thus, the relevant functions for the map $\mathcal{F}$ read $g_1(\eta) = \w_1-\e\sin(\eta)-\sigma\sin(2\eta)$, $g_2(\eta)=\w_2+\e\sin(\eta)+\beta\cos(3\eta) $ and $f(\eta)=g_1(\eta) - g_2(\eta)$. For the chosen parameters, the system attains a stable phase difference $\eta^* \approx -0.12 \pi$. Thus frequency, Floquet exponent, and PRC prefactor follow as $\w \approx 1.01$, $\kappa \approx -0.11$, and $\gamma \approx 0.30$. 

We perform the stimulation experiment by monitoring either the first or the second oscillator. The results are depicted in Fig.~\ref{fig:bifurcation_diagram_osc1} and Fig.~\ref{fig:bifurcation_diagram_osc2}, respectively. In both cases, the proposed theory for the iterated mapping $\mathcal{F}$ corresponds to the direct simulation with Dirac kicks to a large extent. Both agree with the direct simulation by kicks of finite duration $\Delta$ for small $q$; however, the results differ for large $|q|$. This discrepancy is due to the difference in the effect of stimulating with the amplitude $I$ for time $\Delta$ starting at $\vp_0$,  compared to an instantaneous shift of $I\Delta Z(\vp_0)$.

We remark, that since we define the phase $\Phi$ on the limit cycle  as $\Phi = \vp_1$, we have $\Phi_0 = \vp_0$ if the first oscillator triggers the stimulation at $\vp_1 = \vp_0$ and $\Phi_0 = \vp_0 - \eta^*$ if the second oscillator triggers it at $\vp_2 = \vp_0$.

In the first numerical experiment, we monitor the first oscillator. We choose $\Phi_0 = \vp_0 = 3\pi/2$ as the trigger phase since it corresponds to an extremum of $Z$. We observe the appearance of phase slips for $q \gtrsim 0.5$. Since we do not observe phase slips for equally strong negative pulses, we conclude the favorable polarity of the phase shift to be negative (positive kicks at negative PRC value $Z(\vp_0) < 0$). This conclusion corresponds to our choice $\w_1 < \w_2$.

Monitoring the second oscillator, we experiment with two different trigger phases $\Phi_0 = 3\pi/2$ ($\vp_0=\Phi_0+\eta^* \approx 1.62\pi $) and $\Phi_0 = 1.88\pi$ ($\vp_0 \approx 2\pi $). Even though $\Phi_0 = 3\pi/2$ yields an extremum of $Z$, we do not observe phase slips in the shown range of kick actions $q$, neither for positive nor for negative kicks, see Fig.~\ref{fig:bifurcation_diagram_osc2}(a). However, for the value $\Phi_0 = 1.88\pi$, we observe the appearance of phase slips in an interval of $q$. For finite-sized kicks of $\Delta= 10^{-5}$, phase slips occur for $0.49 \lesssim q \lesssim 0.71$. For the Dirac kicks, both for the mapping and the direct simulation, the interval of phase slips is narrower: it starts at $q \gtrsim 0.53$ and ends at $q \lesssim 0.56$. For sufficiently large $q$, a new fixed point is formed. This happens due to the dependence of the kick-induced phase shift on the phase difference $\eta$. In contrast to the case of monitoring the first oscillator, here, the kick-induced phase shift can change its sign depending on the phase difference $\eta_n$. The kick is strong enough for the first few iterations to bring the system out of its potential well. As the system then tends to relax to the next equilibrium value $\eta^*-2\pi$ and reaches the next trigger point $\vp_2 = \vp_0 + 2\pi$, the kick acts in the opposite direction and brings the system up the potential wall again. In this way, the system gets trapped, and a fixed point establishes. For practical purposes of avoiding that scenario, we mention the possibility of pausing the stimulation after one phase slip or varying the kick strength randomly.

Such behavior is not possible if we monitor the first oscillator, at least if there exists only one stable phase difference $\eta^*$ of the unperturbed coupled system: Since the kick-induced phase shift does not depend on the phase difference $\eta$ (at least for Dirac kicks), and thus is constant for a given trigger phase $\vp_0$, it will constantly shift the phase difference in the same direction (the evoked phase shift $A=\const$). Thus, if the kicks are strong enough to induce a phase slip once, they will continue causing them.

\subsection{More than two oscillators: an outlook}
We stress that our proposed strategy of phase-specific pulse stimulation with an observation of the partial periods is generally extendable to systems of more than two coupled units. 
As a particular showcase, we consider a set-up of five globally diffusively coupled Rayleigh oscillators
\begin{align}
    \ddot x_i & - \mu(1-\dot x_i^2)\dot x_i+\w_i^2 x_i
    = \frac{\e}{5} \sum_{k=1}^5 (\dot x_k - \dot x_i) + \delta_{ij} p(t)
    \,,
\end{align}
where $\mu=2.0$, $\e = 0.05$, $\w_1 = 0.99$, $\w_2 = 0.995$, $\w_3 = 1$, $\w_4 = 1.005$, $\w_5 = 1.01$, and $i = 1, \dots, 5$.
The asymptotic autonomous state is the state of global frequency locking. Then, stimulation enters the equation for oscillator $j$. For simplicity, we consider the case of stimulating and observing the same oscillator. Thus, a pulse with the amplitude $q/\Delta$ and duration $\Delta=0.01$ is applied to the system when $x_j=-0.8$ and $\frac{\dd}{\dd t}x_j<0$ (with a ``dead'' time interval of $2.0$ to exclude another pulse within that interval). This threshold-crossing event corresponds to a phase where the PRC of a single Rayleigh oscillator is positive; see Fig.~\ref{fig:PRC comparison}. We observe the emergence of phase slips for three of four tested scenarios: When stimulating the slowest oscillator ($j=1$) with $q \lesssim -0.276$ or $q \gtrsim 0.208$, we achieve the desynchronization of this oscillator from the rest (cluster formation $1:4$). Also, when we stimulate oscillator $j=3$ in the center of the frequency distribution with $q \gtrsim 0.208$, we get a cluster formation of $1:4$, desynchronizing oscillator $3$ from the rest. However, when we stimulate oscillator $3$ with negative pulses, we do not see phase slips for weak pulses (at least for $q>-3$).
The transition from global frequency locking to a quasi-periodic regime with the stimulated oscillator being desynchronized from the rest is observed in all cases. Similar to the case of only two oscillators, it is visible in the partial periods $\tau_n$ of the observed oscillator as a transition from a fixed point to an oscillating pattern. However, we expect that for different frequency distributions, it is also possible to observe mutually desynchronized clusters, i.e., to desynchronize the stimulated oscillator only from a fraction of the population.

Of course, this model is only a particular example of a network of more than two oscillators. In general, one can imagine a coupled oscillator population, where stimulation directly affects a subpopulation, and observation is possible on another group. Then, a broad spectrum of cases is possible depending on the intersection of the stimulated and observed oscillator sets. Furthermore, the formation of clusters depends on the frequency of the stimulated oscillators, the frequency distribution, and, of course, the network connectivity. However, we expect that, in most cases, breaking the global frequency locking with the proposed strategy is possible.

\begin{figure}[!h]
    \includegraphics[width=0.48\textwidth]{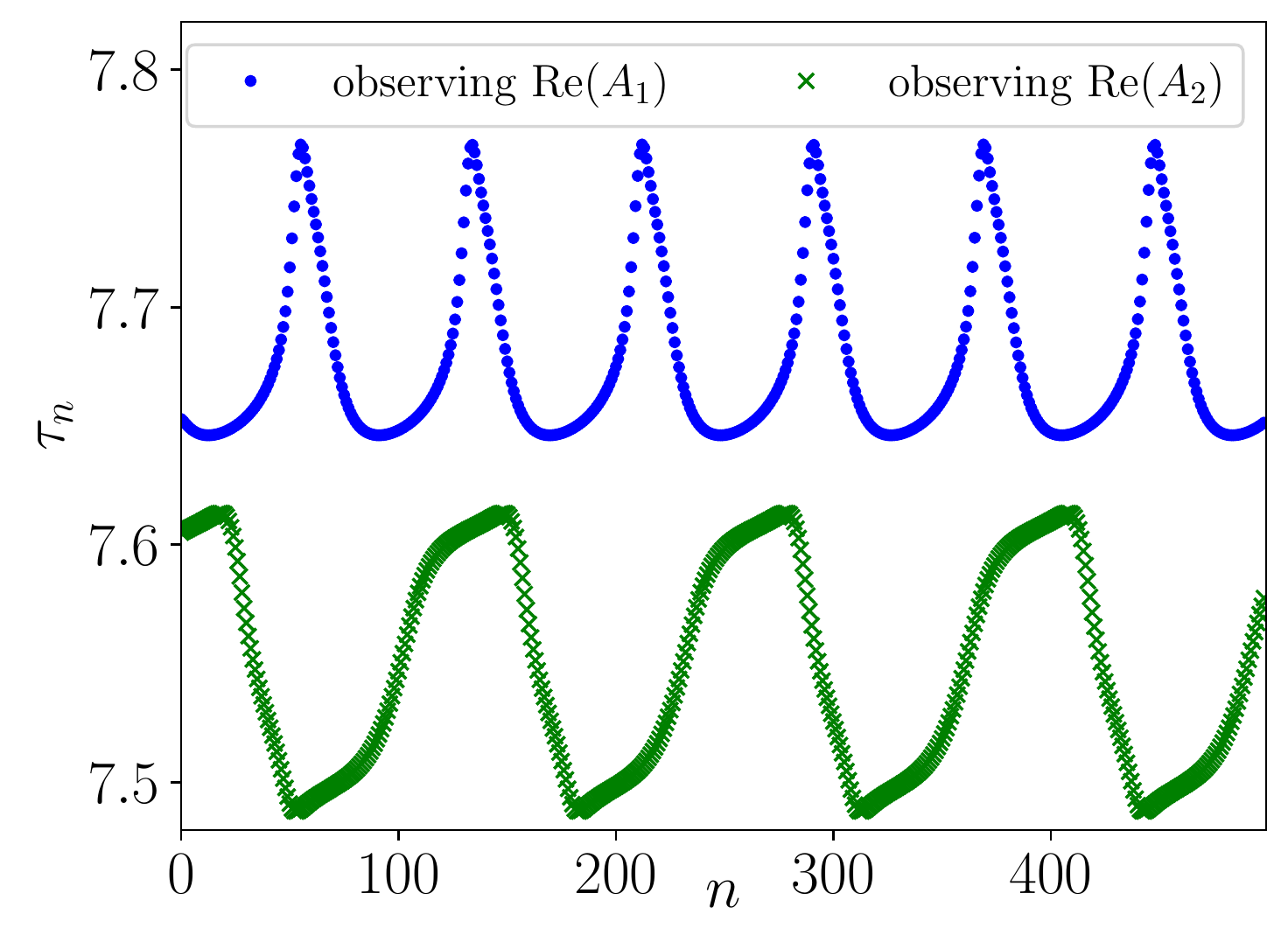}
    \caption{Inter-kick durations (partial periods) $\tau_n$ versus $n$ for two interacting macroscopic oscillators. The observables used are the real parts of the subpopulation mean fields,  $\text{Re}(A_1)$ (blue circles) and $\text{Re}(A_2)$ (green crosses), i.e., the macroscopic oscillations. To demonstrate that the stimulation does not quench the macroscopic oscillations, we compute the minima of $\text{Re}(A_{1,2})$ over the time interval of stimulation. Their values are $\min(|A_1|)\approx 1.017$ and $\min(|A_2|)\approx 1.013$ for the first experiment and $\min(|A_1|)\approx 0.897$ and $\min(|A_2|)\approx 1.014$ for the second experiment, to be compared with the corresponding values $\min(|A_{1,2}|)\approx 1.022$ for the unperturbed system.
    }
    \label{fig:Rayleigh_ensemble}
\end{figure}

To illustrate the applicability of our approach to the case of two interacting macroscopic oscillators, we consider a model of $200$ Rayleigh oscillators grouped into two subpopulations of size $100$ each. 
Each subpopulation is globally coupled; additionally, each unit is coupled to all units of the other subpopulation. (The inter-population coupling is stronger than the intra-population one.) All oscillators of the first subpopulation are subject to stimulation. The model reads
\begin{align}
    \ddot x_i - \mu(1-\dot x_i^2)\dot x_i+\w_i^2 x_i
    &= G_1 + p(t)\,, 
    \quad i \in \llbracket 1, 100 \rrbracket \\
    \ddot x_i - \mu(1-\dot x_i^2)\dot x_i+\w_i^2 x_i
    &= G_2 \;,
    \quad i \in \llbracket 101, 200 \rrbracket
    \,,
\end{align}
where 
\begin{align}
    G_1
    = \frac{\e}{200} \left( \sum_{k=1}^{100} (\dot x_k - \dot x_i) + \alpha \sum_{k=101}^{200} (\dot x_k - \dot x_i) \right)
\end{align}
and
\begin{align}
    G_2
    = \frac{\e}{200} \left( \alpha \sum_{k=1}^{100} (\dot x_k - \dot x_i) + \sum_{k=101}^{200} (\dot x_k - \dot x_i) \right)
    \,.
\end{align}
The parameters of the model are $\mu=2.0$, $\e = 0.2$, and $\alpha = 0.15$. The frequency parameters are normally distributed with the standard deviation $0.005$ and mean values $1.005$ (first subpopulation) and $0.995$ (second subpopulation). The subpopulation mean fields 
\begin{align}
    A_1 &= \frac{1}{100} \sum_{k=1}^{100} \big (x_k + \ii y_k \big )\,, \\
    A_2 &= \frac{1}{100} \sum_{k=101}^{200} \big (x_k + \ii y_k \big ) \,,
\end{align}
quantify the degree of synchrony within each subpopulation and $\text{Re}(A_{1,2})$ represent our observables. 
Without stimulation, the system evolves to a globally synchronous state, i.e., all oscillators are frequency-locked. In this state, we administer pulses each time $\text{Re}(A_1) = -0.8$ and $\frac{\dd}{\dd t}\text{Re}(A_1)<0$ with $q=-0.12$ (first test) and each time $\text{Re}(A_2) = -0.1$ and $\frac{\dd}{\dd t}\text{Re}(A_2)<0$ with $q=0.12$ (second test). 
The stimulus's shape and dead time are identical to the experiment with five oscillators. The results are shown in Fig.~\ref{fig:Rayleigh_ensemble}. The pulses are strong enough to break the inter-population frequency locking, but the subpopulations do not stop oscillating. Similarly to the previous examples of two coupled oscillators, the partial periods $\tau_n$ change from a fixed point behavior to an oscillating pattern as one increases $|q|$ beyond a critical value. Thus, the suggested criterion works for macroscopic oscillators as well.

\section{Finding the proper phase for stimulation}
\label{sec:response}

In the previous section, we have shown that sufficiently intense pulses can induce phase slips if delivered consecutively each time the monitored oscillator attains a pre-selected target phase $\vp_0$. However, the critical kick strength of these pulses to achieve phase slips depends on $\vp_0$, and for some disadvantageous $\vp_0$, it might not work at all. This section illustrates the determination of a proper target phase for that stimulation protocol, which leads to phase slips for as weak pulses as possible.

\subsection{Phase and isostable response curves}

Following Section~\ref{sec:theory}, we consider the dynamics of the synchronized system as a limit-cycle oscillation.
 Correspondingly, this oscillation can be characterized by the phase response curve (PRC) $\mathcal{Z}$. In general, this curve differs from the PRC of the uncoupled oscillator, i.e., $\mathcal{Z}\neq Z$. 
 $\mathcal{Z}$ contains information on how external stimulation shifts the phase of the synchronous oscillation $\Phi$.
Next, the deviation from the limit cycle of the synchronized system is quantified by the isostable response curve (IRC) $\mathcal{I}$~\footnote{Similarly to PRC, the IRC $\mathcal{I}$ is defined as the infinitesimal response in the isostable variable $\psi$ on the limit cycle. It is a function of phase and enters the equation for the isostable variable as $\dot{\psi} = \kappa \psi + \mathcal{I}(\Phi)p(t)$.}.
As discussed in Section~\ref{sec:theory}, the deviation is $\eta-\eta^*$, i.e., it corresponds to the deviation of the phase difference $\eta=\vp_1-\vp_2$ from its stable value. 
The description in terms of PRC and IRC is valid if the system is on or very close to the limit cycle when stimulated. For a detailed explanation, see \cite{wilson2016, wilson2018}.

We derive the PRC $\mathcal{Z}$ from the gradient of $\Phi$ and the PRC of the uncoupled oscillators, both evaluated at the limit cycle:
\begin{equation}
    {\cal Z}(\Phi)
    =\left( \partial_{\vp_i}\Phi \cdot \delta_{i1}Z(\vp_1) + \partial_{\eta}\Phi \cdot Z(\vp_1)
    \right)|_{\eta=\eta^*}
    \,.
\end{equation}
With the partial derivatives $\partial_{\vp_i} \Phi|_{\eta=\eta^*} = \delta_{i1}$ and $\partial_\eta \Phi|_{\eta=\eta^*} = - g_i'(\eta^*)/f'(\eta^*)$, see Eq.~\eqref{eq:phase_general}, we conclude
\begin{align}
     {\cal Z}(\Phi)  = \gamma Z(\Phi) \,.
     \label{eq:relation_PRC}
\end{align}
Thus, the PRC $\mathcal{Z}$ generally differs from the response curve of the first oscillator $Z$ by a factor of $\gamma$. This factor $\gamma$ is characteristic of the coupled system and can potentially take any real value, including $0$. 
Similarly, we derive the IRC by 
\begin{equation}
    {\cal I}(\Phi)
    =\left( \partial_{\vp_i}\psi \cdot \delta_{i1} Z(\vp_i) + \partial_{\eta}\psi \cdot Z(\vp_1) \right) |_{\eta=\eta^*}
    \;.
\end{equation}
Here, the partial derivative with respect to $\vp_i$ vanishes ($\partial_{\vp_i}\psi = 0$) since the isostable variable $\psi$ depends on the phase difference $\eta$ only. The partial derivative with respect to the phase difference $\partial_{\eta}\psi|_{\eta=\eta^*}$ is some constant that depends on the chosen scaling of the isostable variable. Thus, the IRC is proportional to the response curve $Z$ and thus also to $\mathcal{Z}$:
\begin{align}
     {\cal I}(\Phi) \propto Z(\Phi) \propto \mathcal{Z}(\Phi)\;.
     \label{eq:relation_PRC_IRC}
\end{align}
To desynchronize the two oscillators, we want to push their phase difference $\eta$ as far away from its value $\eta^*$ in the locked state as possible. Hence, we want to maximize the response in the isostable variable $\psi$, which is achieved by stimulating the system at a phase that maximizes the IRC $\mathcal{I}$. By relation~\eqref{eq:relation_PRC_IRC}, we have to look for the extrema of $Z$ or $\mathcal{Z}$ to obtain the extrema of the IRC. In the following part of this Section, we will discuss the practical aspects of PRC inference.

\subsection{PRC inference for coupled Rayleigh oscillators}

\begin{figure}[!ht]
    \includegraphics[width=0.48\textwidth]{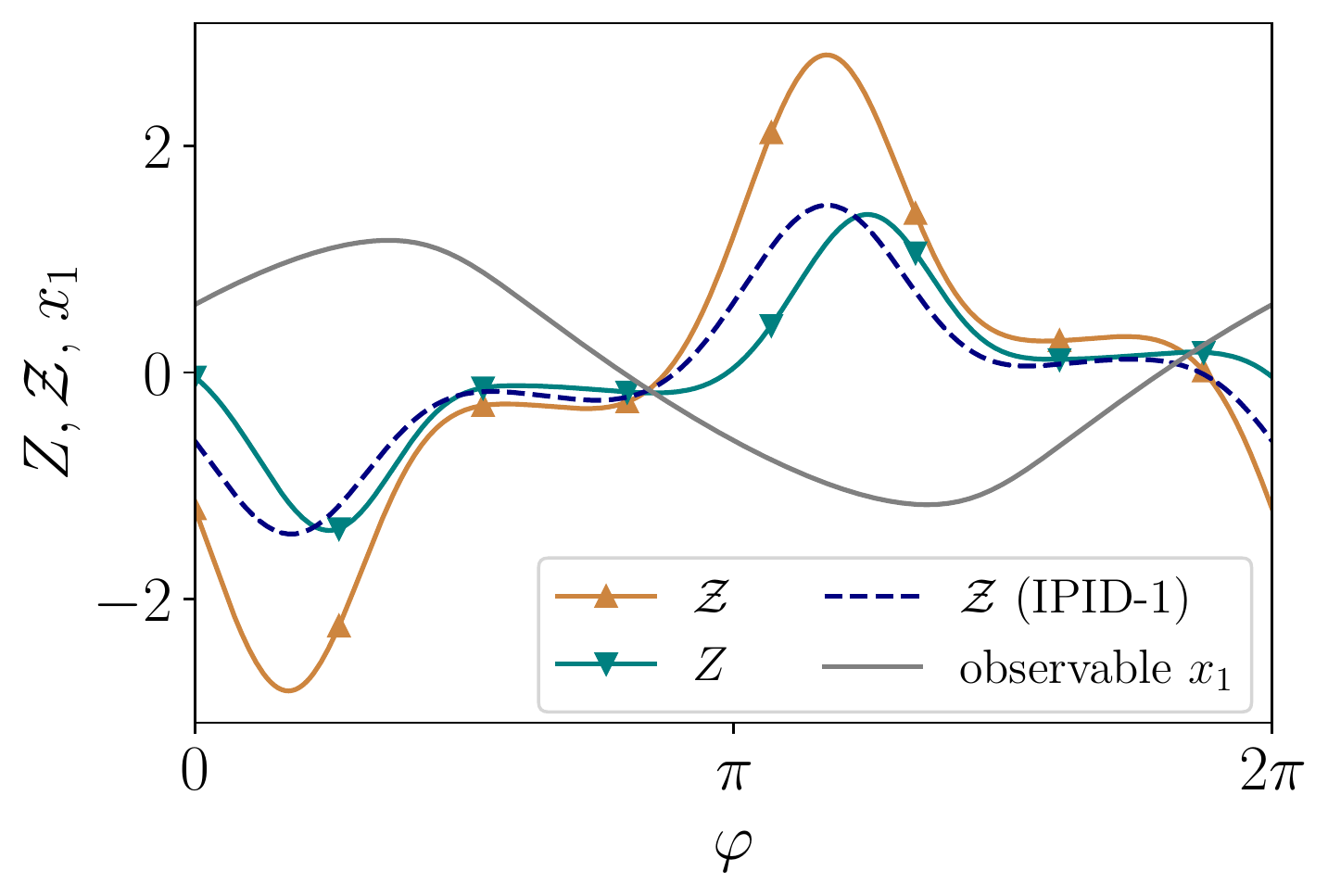}
    \caption{Comparison of PRC inference techniques for the system of coupled Rayleigh oscillators~\eqref{eq:coupled_Rayleighs}, the parameters remain as specified in Sec.~\ref{Sec:II}. The solid orange curve with upper triangles and the dashed purple curve, respectively, depict the resulting PRCs $\mathcal{Z}$ from the standard technique and the IPID-1 method. For comparison, the teal curve with lower triangles illustrates the PRC of the first Rayleigh oscillator $Z$ for the uncoupled case $\varepsilon=0$, obtained by the standard method. In contrast to the relation~\eqref{eq:relation_PRC} between the PRCs of the coupled and uncoupled system in the phase oscillator model, the curves differ not only in scale but also are slightly shifted. The IPID-1-inferred PRC for the coupled system correctly reproduces the PRC's shape but not the scaling. However, the latter is not important for our approach. The solid gray curve is the observable $x_1$ of the coupled system; it provides a map to translate threshold crossings into phases.}
    \label{fig:PRC comparison}
\end{figure}

To demonstrate the PRC inference for the system of two coupled Rayleigh oscillators~\eqref{eq:coupled_Rayleighs}, examined in Section~\ref{Sec:II}, we choose the observable $x_1$ and assign phase values from $0$ to $2\pi$ to one period of the unperturbed oscillation, mapping threshold values of $x_1$ to phases. Thus, instead of operating with phases, we can use the signal values; see the solid gray line in Fig.~\ref{fig:PRC comparison}.

As a benchmark, we exploit the standard approach and apply consecutively single pulses at different phases $\vp$ (i.e., at different signal thresholds) and wait until the system returns to the same state for the $k$-th time; we denote this time interval as $T_k$. Since we are dealing with a weakly stable system~\footnote{For the chosen parameter value $\mu = 2$, individual oscillators are strongly stable, but the limit cycle of the coupled system is weakly stable.}, it may be necessary to wait several periods to ensure that the system has relaxed back to the limit cycle sufficiently close. The PRC then computes as
\begin{align}
    \mathcal{Z}(\vp) = \frac{2\pi}{q} \frac{kT_0 - T_k}{T_0}
    \label{eq:PRC_standard}
    \,,
\end{align}
where $T_0=2\pi/\omega$ is the natural period of the system, and $q$ is the action of the pulse.

A more practical way to infer the PRC is to exploit the newly developed IPID-1 technique~\cite{cestnik2018, cestnik2022}. This technique uses the observed scalar time series and known pulsatile external stimulation to infer PRC via a direct fit of the Winfree equation. See~\cite{cestnik2022b} for the code of implementation.

The standard technique requires at least $k \cdot m$ periods of the oscillation to obtain $m$ data points of the PRC. For example, to compute the PRC via the standard technique in Fig.~\ref{fig:PRC comparison}, we used $k=20$.
IPID-1 needs a substantially shorter observation time to conveniently depict the entire PRC due to the least squares fit. In addition, IPID-1 does not rely on a specially designed stimulation protocol that hits a certain target phase. For example, adding a Poissonian process to the stimulation period suffices. The requirement for IPID-1 is that the time series of both an observable of the system and the external stimulation are known.
The results of the inferred PRC using the standard and IPID-1 methods are depicted and compared in Fig.~\ref{fig:PRC comparison}.

In the more difficult case of observing the second oscillator (which is not directly stimulated), the IPID-1 method failed, for our example, yielding a vanishing PRC. However, the standard method is still applicable in that case.

\subsection{Are the PRCs extrema optimal targets for phase-triggered stimulation?}

Let us assume that we obtained the exact PRC $\mathcal{Z}$ and thus have perfect knowledge about phases (i.e., thresholds) at which the system is displaced most efficiently from the limit cycle. Does that mean we have found the best phase to trigger external pulses?

In the case of monitoring the first oscillator, it indeed does. As we have seen in Sec.~\ref{sec:theory}, the evoked phase shift is constant if the stimuli are applied at the same $\vp_1$ every time. Moreover, selecting the extrema of $\mathcal{Z}$ ensures the maximal phase shift. What remains to be determined is whether the kicks shall be positive or negative, i.e., whether advancing or delaying the system is more efficient in causing phase slips. For our choice, $\w_1 < \w_2$, slowing the first oscillator by negative phase shifts was the favorable choice. For the opposite case, it would be vice versa. We remark that Fig.~\ref{fig:ray_effect} shows the coupled Rayleigh system for a phase-specific stimulation each time $x_1$ crosses the threshold $x_0=1$ from below. This threshold corresponds to a phase of $\vp_1 \approx 0.18\pi$, see Fig.~\ref{fig:PRC comparison}, and is close to the minimum of $\mathcal{Z}$. Thus, it is an excellent choice to induce phase slips for comparably small positive kick actions $q$.

The opposite case of monitoring the second oscillator is more involved. The reason is that the induced phase shifts following a pulse are not constant as in the previous case. By selecting a trigger phase $\vp_2 = \vp_0$, the kick-induced phase shift also depends on the phase difference $\eta$, see Sec.~\ref{sec:theory_osc2}. Thus, even when $\vp_0$ is most effective on the limit cycle at $\eta^*$, it might lose this efficiency for the new phase difference $\hat{\eta}$ that establishes as a result of the consecutive kicks. We do not yet see a practical way to overcome this issue just by knowing the PRC. It might still be a good idea to start exploring efficient phases close to the extrema of $\mathcal{Z}$ since these at least guarantee the most significant possible displacement from the limit cycle for the first few kicks. To avoid a trapping scenario as described in Sec.~\ref{sec:theory_example} and shown in Fig.~\ref{fig:bifurcation_diagram_osc2} we mention the possibility to add a stochastic process to the pulse action $q$.

\section{Discussion}
\label{sec:discussion}

In this article, we have demonstrated how a system of two synchronized oscillators can be desynchronized by short pulses applied to only one of both in a phase-specific manner. We focused on the restriction of having access to the observation of only one of the two units. Otherwise, when signals from two oscillators are available, well-known measures such as the time-averaged phase differences or difference of averaged frequencies will quantify the degree of phase and frequency locking, and tracing desynchronization is trivial.

For both cases of observing the stimulated and the unstimulated oscillator, we showed the efficiency of this approach for a well-chosen trigger phase. We developed a theoretical framework for the approximation of weakly coupled phase oscillators. This framework allowed us to derive an exact expression for the phase of the coupled system. We used it to establish a relation between the coupled system's phase response and the individual oscillator's phase response curve. This relation can be used to find efficient trigger phases for a phase-specific stimulation protocol.
The proposed strategy of phase-specific pulse stimulation is robust to natural frequencies or coupling parameters as long as the assumption of weakly coupled phase oscillators is still applicable.
However, the stimulation efficiency essentially depends on the phase response curve. If the interval of sensitive phases is very narrow, the technique may become less efficient due to imprecision in the phase measurement or will require stronger stimulation. 
In other words, if the response curves' extrema are very narrow, the pulse can hit at the less effective phase, and the kick action has to compensate for that.

In our paper, we treated a deterministic case. Now, we remark on the effect of noise that is two-fold. First, it is well-known that synchronization in the presence of noise is imperfect due to noise-induced phase slips. On the other hand, the real-time phase estimation required for phase-specific pulses becomes imprecise at higher noise levels. Thus, strong noise will result in a non-optimal delivery of pulses but will reduce the level of synchrony by itself.

In particular, we discuss the optimization of the stimulation. The first issue is the polarity of the pulse's action, which determines whether an induced phase shift advances or delays the phase of the stimulated oscillator. We know that phase delays are favorable if the stimulated oscillator is slower than the unstimulated one in the absence of coupling (which is the case in the examples in this article). Vice versa, if it were faster, phase advances would be favorable. However, the induced phase shift is the product of both action and phase response at the trigger phase. Thus, the same pulse can cause advancing and delaying shifts if delivered at different phases. Hence, to account for that consideration, knowledge about the PRC up to a positive factor is also required. We remark that to determine which direction is favorable for the induced phase shift, it must be known whether the stimulated oscillator is faster or slower than the other. Since that is unknown {\it a priori}, we suggest testing both polarities for a given phase with a high phase response in absolute value.

Another issue is how to minimize the number of pulses required to induce a phase slip. We remind that evoked phase slip means that the system escapes the basin of the locally stable phase difference and then evolves toward the next potential minimum, see Fig. 1. Obviously, having reached the local maximum of the potential, the system tends to the next equilibrium state by itself. It does not need additional pulses driving it in that direction. Thus, pausing stimulation after passing the maximum excludes unnecessary intervention and also avoids a trapping scenario described in Section~\ref{sec:response} for the case of monitoring the unstimulated oscillator.
The underlying problem is to detect the instant of passing over the barrier. While the emergence of an oscillating pattern for $\tau_n$ unambiguously reveals phase slips, see, e.g., Fig.~\ref{fig:ray_effect}a and Fig.~\ref{fig:bifurcation_diagram_osc1}b, we do not know how to detect the barrier crossing from this pattern precisely. This task remains an open problem for future research.

To highlight the efficacy of our approach, we compared the phase-specific stimulation to Poisson-distributed inter-pulse intervals with similar statistics. A Poisson-distributed random variable was scaled and shifted to have the same minimal time and expectation value as the partial periods recorded from the phase-specific run. Also, the pulse shape was the same, and the first pulse was applied at the same instant as the phase-specific one. The frequency and time-averaged phase differences indicated that the phase-specific stimulation strategy outperformed the random stimulation with different standard deviations (lower, equal, and larger than the phase-specific stimulation). We expect this result to be robust for other distributions of random inter-pulse intervals. We are confident that our method is superior to randomly delivered kicks with comparable external intervention.

In the following, we comment on the limitations of the theoretical description of our approach. Our considerations rely on weak-coupling approximation with Dirac pulse stimulation. Thus, strongly coupled systems can differ from the phase description used here. Also, the effects of very strong or long stimuli might not be accurately described by the derived dynamical map.

Within our theoretical framework, several questions remain unanswered. First, we do not see a straightforward data-driven way to predict the critical action to induce phase slips. If the dynamical equations are known, the critical action can be found by numerically solving Eqs.~\eqref{eq:map_definition},~\eqref{eq:map_definition_osc2} for a fixpoint as a function of action $q$. The boundaries of existence then mark the critical actions. We rely on continuously increasing the action for unknown dynamical equations until phase slips appear.

Another issue is the optimal trigger phase for the case of monitoring the unstimulated oscillator. As outlined in Section~\ref{sec:response}, it is not necessarily an extremum of the phase response curve that leads to phase slips at all, let alone in the most efficient way. We do not yet see a practical solution apart from trying out different phases in the vicinity of an extremum of the phase response.

Another assumption we made throughout this article was the uniqueness of the system's stable phase difference equilibria. In principle, multiple stable equilibrium states are possible, corresponding to multiple local minima of the potential in Fig.~\ref{fig:potential}. We will discuss such a case now. Unlike the case of a unique stable state, where the system reenters the basin of attraction if the unstable equilibrium is crossed, the system finds itself in the basin of attraction of another stable phase difference. Thus, system quantities like the frequency, the Floquet exponent, and the PRC scaling factor $\gamma$ can change. The individual phase response $Z$ remains constant, though. This new basin might be impossible to leave with the same kicks that kicked it there in the first place. If we monitor the stimulated oscillator, increasing the kick action will eventually suffice to leave the basin. Repeating this procedure for potentially more stable states will result in a kick action large enough to leave all basins and thus induce phase slips: a repeating visit of all basins. There is no such guarantee for monitoring the unstimulated oscillator, and we cannot exclude that it might be necessary to change the trigger phase depending on the current basin.

\begin{acknowledgments}
E.T.K.M. acknowledges financial support from Deutsche Forschungsgemeinschaft (DFG, German Research Foundation), Project-ID 424778381 – TRR 295. We thank Prof. A.~Gharabaghi for inspiring discussions.
\end{acknowledgments}

\nocite{*}

\end{document}